\documentclass[doublecol]{epl2} 

\usepackage{amssymb}
\usepackage{amsmath}
\usepackage{graphicx}
\usepackage{dcolumn}
\usepackage{bm}
\usepackage{url}

\newcommand{\ud}{\mathrm{d}}
\newcommand{\Sd}{\frac{2\pi^{\frac{d}{2}}}{\Gamma\left(\frac{d}{2}\right)}}

\title{An Information Theoretic Approach to Statistical Dependence: Copula Information}
\shorttitle{Information Theory and Copulas}

\author{R.S. Calsaverini\inst{1} \and R. Vicente\inst{2}}
\shortauthor{R.S. Calsaverini and R. Vicente}

\institute{                    
  \inst{1} Dep. de F\'{\i}sica Geral, Instituto de F\'{\i}sica, Universidade de S\~ao Paulo, CP 66318, 05315-970, S\~ao Paulo-SP, Brazil\\
  \inst{2} {Complex Systems}, Escola de Artes, Ci\^encias e Humanidades, Universidade de S\~ao Paulo, 03828-020, S\~ao Paulo-SP, Brazil
}

\abstract{
We discuss the connection between information and copula theories by showing  that a copula can be employed to decompose the information content of a multivariate distribution into marginal and dependence components, with the latter quantified by the mutual information. We define the information excess as a measure of deviation from a maximum entropy distribution. The idea of marginal invariant dependence measures is also discussed and used to show that empirical linear correlation underestimates the amplitude of the actual correlation in the case of non-Gaussian marginals. The mutual information is shown to provide an upper bound for the asymptotic  empirical log-likelihood of a copula. An analytical expression for the information excess of T-copulas is provided, allowing for simple  model identification within this family.  We illustrate the framework in a financial data set.}

\begin{document}
\maketitle

\section{\label{sec:intro}Introduction}
Modeling statistical dependence has a pervasive role in science. Information theory provides a unifying framework for ideas from areas as diverse as differential geometry \cite{amari}, physics \cite{ariel,kanter,maccone}, statistics and telecommunications \cite{cover}. From the information theoretic point of view dependence can be quantified by measuring the distance between a given model defined by a joint probability density $\phi(\mathbf{x})$ and a mean field model defined by 
$\phi_0=\prod_{j=1}^N f_j(x_j)$, where $f_j(x_j)$ are marginal densities $f_{j}(x_j)=\int \prod_{k\neq j} dx_k\; \phi(\mathbf{x})$ \cite{meanfield}. The relative entropy given by 
\begin{equation}
 S\left[\phi \mid\mid \phi_0\right]=\int \prod_{j=1}^N dx_j\, \phi(\mathbf{x}) \log \left( \frac{\phi(\mathbf{x})}{\prod_{j=1}^N f_j(x_j)} \right).
\end{equation}
defines a premetric in the space of distributions that can be employed to quantify the degree of dependence in a model, this particular measure is also known as the total correlation or, in the bivariate case, as the mutual information.

The copula theory has been proposed in statistics as an approach for modeling general dependences in multivariate data. 
A theorem due to Sklar \cite{nelsen} assures that, under very general conditions, for any joint cumulative distribution function (CDF) $F(\mathbf{x})=\prod_{j=1}^N\int_{-\infty}^{x_j}  dx_j\, \phi(\mathbf{x})$ there is a function $C(\mathbf{u})$ (known as the \textit{copula function}) such that the joint CDF can be written as a function of the marginal CDFs in the form $F(\mathbf{x})=C[F_1(x_1),\cdots F_N(x_N)]$. The converse is also true: this function couples any set of marginal CDFs to form a multivariate CDF. This provides a convenient picture of the marginals as being responsible for the idiossincratic properties of each variable and the copula function as a description of the dependence between them.  

A complete articulation of these two concepts is, however, curiously absent in the literature. In this short contribution we seek to survey the basic ideas connecting these two threads emphasizing the information theoretic interpretation. 

We have organized this letter as follows. In the next section we briefly discuss the idea of  measures of dependence that are marginal invariant. We then connect copula theory with mutual information by introducing the concept of copula information and   present an analytical prescription to identify a model for bivariate non-Gaussian dependences within the T-copula family by estimating the mutual information. We briefly comment on general consequences and perspectives in a final section.

\section{\label{sec_mutual} Mutual information and copulas}
From this point on we restrict our discussion to  bivariate distributions, the multivariate case follows after  straightforward adaptations. 

Two random variables $X$ and $Y$ are said to be statistically dependent if, and only if, their joint probability density function (PDF) cannot be written as a product of marginal PDFs, that is, if $\phi(x,y)\neq f_{x}(x)f_{y}(y)$, where $f_{x}(x)$ and $f_{y}(y)$ are marginal densities. A convenient way to quantify statistical dependencies is by evaluating the mutual information defined by:
\begin{equation}
I(X,Y)=\int dx dy\, \phi(x,y) \log \left( \frac{\phi(x,y)}{f_x(x) f_y(y)} \right).
\label{eq:IM}
\end{equation}
This quantity is a premetric, to say, it is positive and only vanishes in the case of independent variables. By defining the entropy of the distribution of $X$ as $S[f_x] = \int dx\, f_x(x) \log f_x(x)$ and the average conditional entropy as $S[f_{x|y}] = \int dy\, f_y(y)\,\int dx\, f_{x|y}(x) \log f_{x|y}(x)$, where   $f_{x|y}(x)$ denotes the conditional probability of $X$ given $Y$, the identity  
\begin{equation}
I(X,Y) = S[f_x] - S[f_{x|y}]
\end{equation}
 provides an interpretation for the mutual information as the average reduction in the uncertainty in $X$ given knowledge of $Y$. Alternatively, the mutual information can be regarded as a distance to statistical independence in the space of distributions measured by the relative entropy  between the actual joint distribution and the product of marginals $I(X,Y)=S[\phi\mid\mid f_x f_y]$.

 Sklar's theorem asserts that there exists a copula function such that the joint CDF can be written as $F(x,y)=C[F_x(x),F_y(y)]$. We may also regard a copula function as the joint CDF of two uniformly distributed variables $u$ and $v$, both in the $[0,1]$ interval. Such a pair $(u,v)$ can always be found from any pair of random variables with the substitution $u = F_{x}(x)$ and $v = F_{y}(y)$.

To exemplify we can build a joint standard Gaussian with correlation $\rho$ by plugging Gaussian marginal distributions $\Phi(x)=\int_{-\infty}^x \frac{du}{\sqrt{2\pi}}\, e^{-\frac{1}{2}u^2}$  into the Gaussian copula defined as:
\begin{equation}
C[u,v]=\Phi_\rho\left(\Phi^{-1}(u),\Phi^{-1}(v)\right),
\end{equation}
where $\Phi_\rho(x,y)=\int_{-\infty}^x\int_{-\infty}^y\frac{du dv}{\sqrt{4\pi^2 (1-\rho^2)}}\, e^{-\frac{u^2+v^2-2uv\rho}{2(1-\rho^2)}}$. 

Clearly $X$ and $Y$ are dependent if, and only if, $C[u,v]\neq uv$. Introducing the copula density as $c[u,v]=\frac{\partial^2}{\partial u \partial v} C[u,v]$, we can  decompose  the joint probability density as 
\begin{equation}
 \phi(x,y)=c[F_x(x),F_y(y)]f_x(x)f_y(y) 
 \label{eq:joint}
\end{equation} 
and observe that statistical dependence would simply imply  that $c[u,v]\neq 1$. 

\section{\label{sec:rank} Marginal invariant measures}
Two close concepts in statistics are dependence and concordance. While dependence relates only to the functional relationship between two variables, concordance measures whether  positive or negative comovement of variables is present. Measures of dependence and concordance are plenty. However,  a good dependence (resp., concordance) measure should \cite{nelsen,kotz}:
\begin{enumerate}
 \item be invariant under reparametrizations: $(x,y) \rightarrow \left(q(x),w(y)\right)$, if $q(x)$ and $w(y)$ are monotonous functions (changing sign if one of the {re\-pa}{\-ra\-me\-tri\-za\-tions} is a monotonically decreasing function, in the case of concordance measures),
 \item have a unique minimum (a unique zero, in the case of concordance), that can be set to zero with no loss of generality, at $\phi(x,y)=f_{x}(x)f_{y}(y)$.
\end{enumerate}	
 Some authors would also require that a measure of dependence (concordance) should be restricted to the $[0,1]$ ($[-1,1]$) interval. We do not require it here since any real number can be trivially mapped into any interval. Good measures of concordance on the other hand must have a unique zero if X and Y are statistically independent, be invariant under monotonically increasing reparametrizations and change sign if one of the functions of the reparametrization is monotonically decreasing. 

With the concept of copula density at hand, these desiderata can be concisely restated as: a measure of dependence must be a functional of the co{\-pu}{\-la} density alone (i.e. must be independent of marginal densities), with a unique minimum at $c[u,v]=1$.

The linear correlation for standardized variables $\rho(X,Y)=\int dx dy \,xy \,\phi(x,y)$ is widely used as a measure of concordance and its absolute value as a measure of dependence. The correlation may be rewritten in terms of copula densities as: 
\begin{equation}
 \rho(X,Y)=\int_{[0,1]^2} du dv \,c[u,v]\, F_x^{-1}(u)F_y^{-1}(v)   
\label{eq:correl1}
\end{equation}
If $X$ and $Y$ are independent, $c[u,v]=1$ and consequently $\rho(X,Y)=0$. However, it is clear that a copula may be chosen such that the linear correlation vanishes even though $c[u,v]\neq 1$. Moreover, $\rho(X,Y)$ is obviously dependent on  marginal distributions.

A better alternative for measuring concordance would be the rank correlation, also known as Spearman's $\rho$ defined as
\begin{equation}
 \rho_{\mbox{rank}}(X,Y)=12\int_{[0,1]^2}du dv\,c[u,v]\, uv  - 3 .
\label{eq:correl2}
\end{equation}
This measure strictly fulfills concordance measures desiderata.  For a Gaussian bivariate distribution, the rank correlation is related to the correlation parameter as:
\begin{equation}
 \rho_{\mbox{rank}}[\Phi_\rho] = \frac{6}{\pi} \sin^{-1}(\frac{\rho}{2}) 
 \label{eq:rankgauss}
\end{equation}
Where $\rho$ is the correlation parameter of the Gaussian copula, which is identical to the usual linear correlation \textit{only if the marginals are also Gaussian}. Another  measure of dependence that is marginal independent is Kendall's tau defined as
\begin{equation}
 \tau(X,Y)=4\int_{[0,1]^2} dC[u,v]\, C[u,v] - 1.
\label{eq:tau}
\end{equation}

In the case of  meta-elliptical distributions \cite{Fang02}, that includes Gaussian and T copulas, Kendall's tau is also related to the the correlation parameter as:
\begin{equation}
 \tau=\frac{2}{\pi}\sin^{-1}\left(\rho\right).
\label{tau_T}
\end{equation}

In the next section we show that the mutual information  also fulfils good dependency measures desiderata, since it  is always non-negative, it only vanishes for independent variables and it is a functional of the copula density alone.

\section{\label{sec:entropy} Copula Information}

Mutual information and  copula densities can be connected by plugging  eq.~(\ref{eq:joint}) into eq.~(\ref{eq:IM}), and by performing the simple change of variables $u = F_{x}(x)$ and $v = F_{y}(y)$, to conclude that:
\begin{equation}
I(X,Y)=\int_{[0,1]^2} du dv \, c[u,v]\log\left( c[u,v]\right) = -S[c], 
\label{eq:entropy}
\end{equation}
where $S[c]$ is the differential entropy associated with the $c[u,v]$ distribution, which we will (following  \cite{Ma08}) conveniently name the \textit{copula entropy}. Notice that $S[c]\leq 0$, as can be shown by considering eq.~(\ref{eq:joint}) together with Jensen's inequality, since  $-\log(x)$ is a convex function.   This simple result shows that  mutual information is invariant under arbitrary choices of marginal densities $f_{x}(x)$ and $f_{y}(y)$. It is also implied by this connection that using a maximum entropy principle to choose a copula function given constraints is analogous to assuming the least informative dependence (minimum mutual information) which explains the constraints, which is actually a reasonable principle \cite{jaynes}. This provides yet another interpretation for  mutual information: it  quantifies the information content of the coupling (copula) functional. From the identity $S[\phi] = S[f_x] + S[f_y] - I(X,Y)$ and eq.~(\ref{eq:entropy}), we have:
\begin{equation} 
	S[\phi] = S[f_x] + S[f_y] + S[c].
	\label{eq:entropydecomposition}
\end{equation} 
In words: the total information content can be \textit{uniquely} decomposed into the information content in each variable plus the information content on the dependence between them.

\section{\label{sec:excess}  Information excess}
When quantifying  dependence, it is a common practice to start by measuring  linear correlation. In the language we have 
 introduced that is analogous to assuming a Gaussian copula described by a single parameter $\rho$.  However the notion that this parameter can be measured by the usual  linear correlation relies upon the additional assumption that marginals are also Gaussian, as the linear correlation is a measure that also depends on marginals. This particular copula is a very special case as it assumes that the  information contained in the dependence between variables is minimal given $\rho$.  This minimal mutual information content in a Gaussian copula is given by \cite{cover}:
\begin{equation}
   I_{\mbox{Gauss}}(\rho)  =  -\frac{1}{2} \log (1-\rho^2)
\label{eq:Igauss}
\end{equation}
which can also be written as a function of the \textit{observable} rank correlation using eq.~(\ref{eq:rankgauss}). If this assumption of minimal dependence given the parameter $\rho$ fails, an excess of information in the dependence with respect to the Gaussian $I_{\mbox{excess}} = I(X,Y) - I_{\mbox{Gauss}}(\rho)$ is observed. An algorithm for efficient estimation of the mutual information $I(X,Y)$ has been  proposed in \cite{Kraskov2004} which, together with a good estimate for $\rho$, provides a diagnostic tool for information excess. The observation of  excess means that the dependence cannot be specified by the linear correlation alone even after the {identification of} non-Gaussian marginals.

\begin{figure}[h]
	\centering
	\includegraphics[width = 0.7\columnwidth]{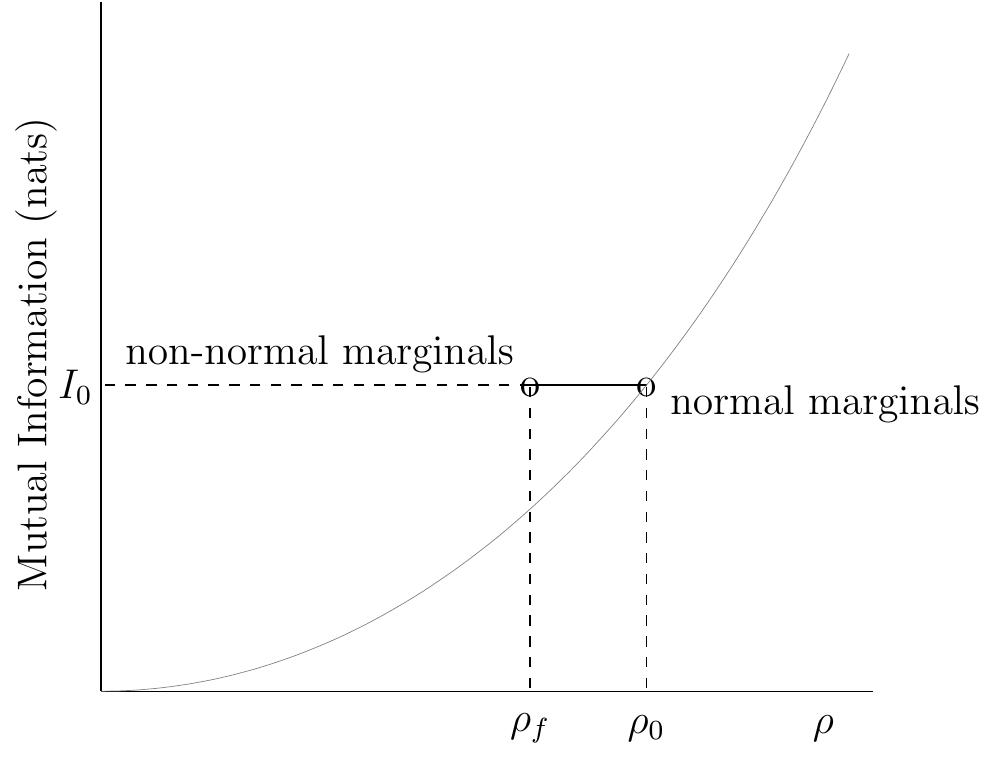}
		\caption{\label{fig:under} \textit{Linear correlation is underestimated} in the case of non-Gaussian marginals. If both marginals and copula are Gaussian the joint distribution can be placed over the lower bound for the mutual information. A change in marginals keeping the copula fixed, preserves the mutual information, however correlation estimates are displaced inwards. } 
\end{figure}

 If marginals are non-Gaussian neither the mutual information  nor the parameter $\rho$  are affected, however, the linear correlation estimate $\rho(X,Y)$ consistently underestimates $|\rho|$. That can be seen by considering the $I(X,Y)$ versus $\rho$ plane in which the curve described by eq.~(\ref{eq:Igauss}) represents a lower bound for the mutual information as depicted in fig.~\ref{fig:under}. For a Gaussian copula the parameter $\rho$ 
is measured by the linear correlation only if marginals are also Gaussian, in this case we can locate a particular joint probability density over the curve of minimal mutual information with a given $\rho$. Suppose that marginals are changed into non-Gaussian densities. As the copula for the variables is unaltered the mutual information is also unchanged, however, the linear correlation can change. As the curve represents a lower bound for the  mutual information given $\rho$, it is only possible for the linear correlation to change inwards, hence underestimating $|\rho|$. In order to find  $\rho$ correctly  we have first to estimate a  measure that is marginal invariant, as the rank correlation given by eq.~(\ref{eq:correl2}), and then employ an inversion relation as eq.~(\ref{eq:rankgauss}).

\begin{figure}[h]
	\centering
	\includegraphics[width = 0.9\columnwidth]{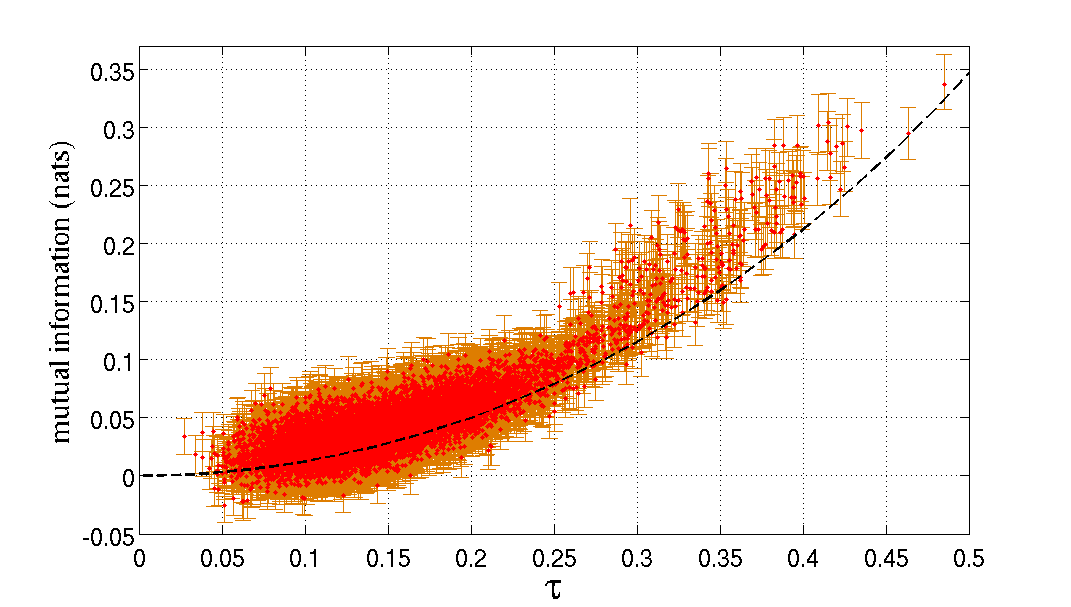}
	\caption{\label{fig:mutualinfostocks} \textit{Mutual Information estimates} following \cite{Kraskov2004} versus Kendall's tau for pairs of series of daily log-returns ($\log[P_{\mbox{close}}/P_{\mbox{open}}]$, where $P_{\mbox{close}}$ and $P_{\mbox{open}}$ are, respectively, close and open prices) of 150 stocks composing the  S{\&}P500 index over the period from January 2, 1990 to September 16, 2008 (around 4700 samples per series). Bootstrap error bars represent a $90\%$ confidence interval. Note that within this confidence interval a great number of the pairs display a non-zero information excess with respect to the {Gaussian} copula}
\end{figure}

As an applied example, fig.~\ref{fig:mutualinfostocks} shows estimates for the mutual information obtained {by the Kraskov-Stogbauer-Grassberger (KSG) method \footnote{We provide a C++ library to calculate Mutual Information with this method with confidence bands estimated with the bootstrap technique \cite{Efron1993} in \url{http://code.google.com/p/libmi/} }} \cite{Kraskov2004}  against Kendall's tau for pairs of series with daily log-returns $r_t=\log[P_{\mbox{close}} / P_{\mbox{open}}]$ (where $P_{\mbox{close}}$ and $P_{\mbox{open}}$ are, respectively, close and open prices) of 150 stocks  composing the Standard \& Poors 500 index (S{\&}P500) over the period from January 2, 1990 to September 16, 2008 (around 4700 points in each series). The error bars have been obtained employing the  bootstrap technique \cite{Efron1993}.  The information excess observed can be traced to time-varying cross-correlations \cite{conlon09} and to dependences between cross-correlations and returns \cite{longin01}  that jointly yield non-Gaussian copulas. Here we have used Kendall's  tau as a marginal invariant measure. In the next section we show that this particular  marginal invariant plane defined by mutual information and Kendall's tau is sufficient to identify the best T-copula representing non-Gaussian dependences shown by the data.

\section{\label{sec:find_copula} Copula Identification}

Given a data set $\{(x_t,y_t)\}_{t=1}^T$ independently sampled from an unknown joint density $\phi(x,y)$, the best approximation $\phi_\theta(x,y)$ within a  manifold  ${\cal F}$,  parameterized by $\theta$, can be found by minimizing a sample estimate of the  relative entropy \cite{meanfield}:
\begin{eqnarray}
 S[\phi\mid\mid\phi_\theta]=\int dxdy\,\phi(x,y)\log\left[\frac{\phi(x,y)}{\phi_\theta(x,y)} \right].
\end{eqnarray} 

By considering eq.~(\ref{eq:joint}) and performing appropriate variable changes we can write:
\begin{eqnarray}
 S[\phi\mid\mid\phi_\theta]= S[c\mid\mid c_{\theta_c}]+S[f_x\mid\mid f_x^{\theta_x}]+S[f_y\mid\mid f_y^{\theta_y}],
\label{relative_entropy}
\end{eqnarray} 
which is just the  decomposition (\ref{eq:entropydecomposition}) in terms of relative entropies. Thus it is reasonably clear that the inference procedure can be implemented by independently minimizing the relative entropy for empirical marginals and copula density. By employing relationship (\ref{eq:entropy}), the contribution from  the copula in eq.~(\ref{relative_entropy}) can be further rewritten as:
\begin{eqnarray}
 S[c\mid\mid c_{\theta_c}]= -L_{\infty}(\theta_c) - I(X,Y) \ge 0,
\label{likelihood}
\end{eqnarray} 
where $L_{\infty}(\theta_c)=\int_{[0,1]^2} du dv\, c[u,v] \log \left(c_{\theta_c}[u,v]\right)$ is the asymptotic copula log-likelihood. Notice that Jensen's inequality implies that $-L_{\infty}(\theta_c)\ge I(X,Y) \ge 0$.  Consequently,  minimizing $S[c\mid\mid c_{\theta_c}]$ is equivalent to maximizing the likelihood with  the mutual information $I(X,Y)$ as a bound. 

\begin{figure}[h]
	\centering
	\includegraphics[width = 0.9\columnwidth]{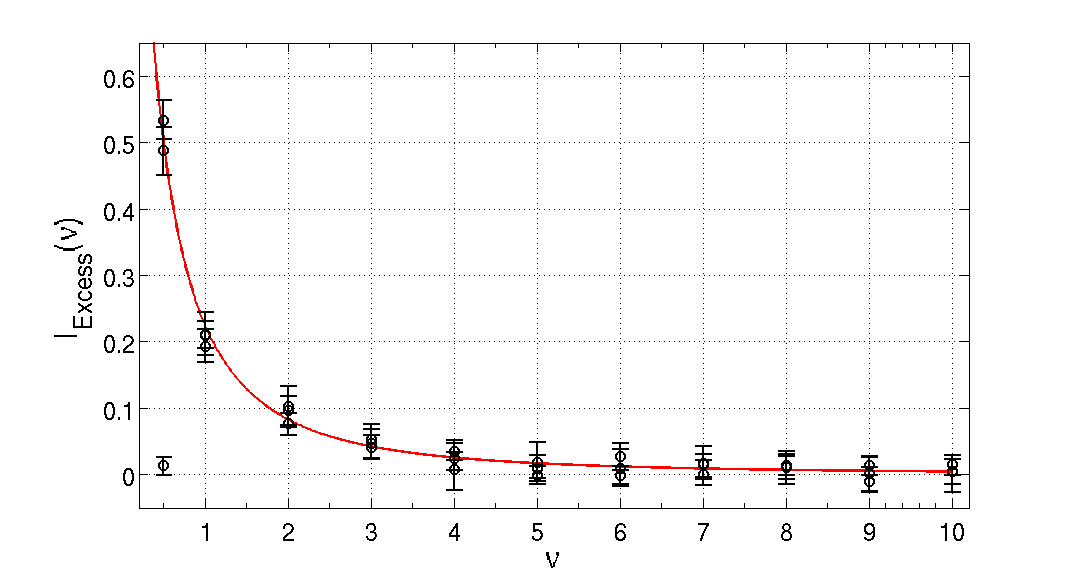}
	\caption{ \textit{T-copula information excess.}  $I_{\mbox{Excess}}(\nu)$ as provided by eq.~ (\ref{iexcess}) (full line). Circles show estimates for 20 runs with T-copulas with known $\nu$ for $\rho=0.1,0.5,0.9$ and arbitrary marginals. Error bars represent one standard deviation. Estimates have been computed by employing {the KSG method}.}
	\label{fig:IExcess}
\end{figure} 

The estimation of $I(X,Y)$  can be employed to measure the quality of a fit within the chosen family ${\cal F}$. In particular, suppose we choose a family such that $L_{\infty}(\theta_c)$ is known analytically. If we additionally find a family that contains a distribution that saturates the bound, we  can use an efficient estimator for the mutual information as \cite{Kraskov2004} to identify the best copula $\theta_c$ within ${\cal F}$ right away. 

In this procedure the identification of the copula is from the start disentangled from the choice of marginals. The T-copula is an interesting choice as the mutual information can be analytically evaluated. The T-copula density is defined in two dimensions as:

\addtocounter{equation}{-1}
\begin{floatequation}
\mbox{\textit{see eq.~\eqref{T-copula}}}
\end{floatequation}

\begin{widetext}
\begin{equation}
	c_{\nu, \rho} [u,v] = \frac{\Gamma(\frac{\nu+2}{2})\Gamma(\frac{\nu}{2})}{\left[\Gamma(\frac{\nu+1}{2})\right]^2{\sqrt{1-\rho^2}}}\frac{\left[1+ \frac{q_{\rho}(t_\nu^{-1}(u),t_\nu^{-1}(v))}{\nu}\right]^{-\frac{\nu+2}{2}}}{\left[1+ \frac{(t_\nu^{-1}(u))^2}{\nu}\right]^{-\frac{\nu+1}{2}} \left[1+ \frac{(t_\nu^{-1}(v))^2}{\nu}\right]^{-\frac{\nu+1}{2}} }
\label{T-copula}
\end{equation}
\end{widetext}

with $q_{\rho}(x,y) = \frac{x^2+y^2-2\rho xy}{1-\rho^2}$ and $t_\nu^{-1}(u)$ denoting the inverse of the distribution function of the univariate Student T density with $\nu$ degrees of freedom.
It can be shown (see appendix) that the mutual information of a multivariate T-copula can be decomposed as:
\begin{equation}
I_{\mbox{T}}(\rho,\nu) =  I_{\mbox{Gauss}}(\rho) + I_{\mbox{Excess}}(\nu),
\end{equation}
where, in two dimensions (2D), $I_{\mbox{Gauss}}(\rho)$ is given by eq.~(\ref{eq:Igauss}). The excess information term only depends on the number of degrees of freedom $\nu$. In 2D it is given by: 

\begin{eqnarray}
 I_{\mbox{Excess}}(\nu) &=& 2\log\left(\sqrt{\frac{\nu}{2\pi}}B\left(\frac{\nu}{2},\frac{1}{2}\right) \right) - \frac{2+\nu}{\nu} \nonumber \\
&+&(1+\nu)\left[ \psi \left(\frac{\nu+1}{2}\right) - \psi \left(\frac{\nu}{2}\right) \right],
\label{iexcess}
\end{eqnarray}

where $B(x,y)$ is the Beta function defined as
\begin{equation}
B(x,y) = \frac{\Gamma(x)\Gamma(y)}{\Gamma(x+y)}
\end{equation}
and $\psi(x)$ is the digamma function. Fig.~\ref{fig:IExcess} shows the T-copula information excess $I_{\mbox{Excess}}(\nu)$ as provided by eq.~(\ref{iexcess}).  The parameter $\rho$ yields the linear correlation in the purely Gaussian case ($\nu\rightarrow\infty$) but must be estimated through a marginal independent measure of concordance/dependence in the general case. For T-copulas  $\rho_{\mbox{rank}}$ is a function of both $\rho$ and $\nu$ that is not known in any simple form. However, in order to identify the appropriate T-copula a simpler alternative is to employ Kendall's tau that is a function of $\rho$ given by eq.~(\ref{tau_T}). We can estimate Kendall's tau and then employ the excess of information in relation to a Gaussian copula to find $\nu$. Fig.~\ref{fig:IExcess}  shows the result of simulations using data sampled from a joint distribution composed by a copula density with known parameters and  arbitrary marginals. Going back to fig.~\ref{fig:mutualinfostocks} the best copula within the manifold of T-copulas can be immediately identified for each point in the mutual information versus Kendall's tau plane.

\section{\label{sec:conclusion} Conclusions}

The literature on information and copula theories has developed in relative isolation. In this paper we sought to discuss a couple of consequences yield by connections between these two threads.

Copula theory can be employed for factorizing a general joint distribution into marginal fluctuations and a dependence core that is not unique. On the other hand, a combination of copula and information theories provides a unique decomposition in terms of global information content measures. This decomposition yields  a simple test of Gaussianity through the estimate of the information excess (a procedure that is simpler than e.g. \cite{Sornette03} or \cite{Panchenko05}) and also suggests a method for copula identification based on  information content matching. This method displays a simple  formal equivalence to the usual maximum likelihood methods (e.g. \cite{Fernandez08}). A C++ library for determining Mutual Information from pairs of time series with the KSG algorithm and bootstrap confidence bands produced by the authors is available at \url{http://code.google.com/p/libmi/}. 

This approach also clarifies the danger of using linear correlation as a measure of dependence for, e.g., portfolio optimization or time series analysis as this measure is bound to underestimate dependence that would be better captured by easily  estimated marginal invariant measures \cite{kotz,Kraskov2004}.Finally, we think that a unified understanding of   information and copula theories may be a useful source of new fundamental ideas for the analysis of multivariate data arising from complex physical phenomena.

\acknowledgments
This work has been funded by Conselho Nacional de Desenvolvimento Cient\'{i}fico e Tecnol\'{o}gico (CNPq) under grant 550981/2007. We thank Nestor Caticha for interesting discussions on maximum entropy methods.

\section{\label{sec:student} Appendix: Information Excess for T Copulas }
\setcounter{equation}{0}
\renewcommand{\theequation}{A.\arabic{equation}}

In this appendix we present a derivation of the entropy and mutual information of  Student T distributions.

The standard Student T distribution in $d$ dimensions is given by:
\begin{equation}
	p_{d} (\mathbf{t}\mid \hat{\Sigma},\nu) =  \frac{1}{Z_d(\hat{\Sigma},\nu)}\left[1+ \frac{\mathbf{t}^{\mathrm{T}}\hat{\Sigma}^{-1}\mathbf{t}}{\nu}\right]^{-\frac{\nu+d}{2}}
\label{standard_studentT}
\end{equation}
where $\nu$ is a parameter and  $\hat{\Sigma}$ is the correlation matrix.

The normalizing prefactor is defined as:
\begin{equation}
\label{eq:partitionfunction}
 \frac{1}{Z_d(\hat{\Sigma},\nu)} =  \frac{\Gamma(\frac{d}{2})}{B(\frac{\nu}{2},\frac{d}{2})\sqrt{(\pi\nu)^{d}|\hat{\Sigma}|}}
\end{equation}
with $B(x,y) = \frac{\Gamma(x)\Gamma(y)}{\Gamma(x+y)}$ being the Beta function.

For $d=2$ this simplifies to read:
\begin{equation}
	p_2(x,y\mid  \rho, \nu ) = \frac{\Gamma(1+\frac{\nu}{2})}{\Gamma(\frac{\nu}{2}){\pi\nu\sqrt{1-\rho^2}}}\left[1+ \frac{q_{\rho}(x,y)}{\nu}\right]^{-(1+\frac{\nu}{2})}
\end{equation}
with $q_{\rho}(x,y) = \frac{x^2+y^2-2\rho xy}{1-\rho^2}$.

The differential entropy of a given set of variables $\mathbf{t}$ distributed as $p(\mathbf{t})$ is given by:
\begin{equation}
	S[p_d] = - \int \ud^n \mathbf{t}\;  p_d(\mathbf{t}) \log\left(p_d(\mathbf{t})\right).
\end{equation}

The mutual information for d dimensions is:
\begin{equation}
 I(X_1,X_2,\ldots, X_d) = \int p_{d}(\mathbf{x}) \log \left[\frac{p_{d}(\mathbf{x})}{\prod_{i=1}^{d} p_{1;i}(x_i)  }\right].
\end{equation}
For variables with identical marginals (for $p_{1;i}(x) = p_1(x)$ for all $i=1,2,\ldots,d$) this can be written in terms of entropies as:
\begin{equation}
\label{eq:multimutual}
 I(X_1,X_2,\ldots,X_d) = d\,S[p_1] - S[p_d]
\end{equation}

We employ a ``replica trick'' \cite{dotsenko} to write:
\begin{equation}
\label{eq:n-entropy}
 	S[p_d] =  - \lim_{n\rightarrow 0} \frac{\ud}{\ud n}\int \ud^n \mathbf{t}\;p_d(\mathbf{t})^{n+1},
\end{equation}
where the limit $n\rightarrow 0$ can be regarded as a analytical continuation for a sequence of integers that is known to give sensible results if $p_d(\mathbf{t})$ has a unique extremum. This calculation is not rigorous but is nicely verified by simulations depicted in fig.~\ref{fig:IExcess}.

We can always simplify this integral by making the transformation $\mathbf{x} = \mathrm{\hat{U}}\mathbf{t}$ where $\mathrm{\hat{U}}$ is the unitary matrix that diagonalizes $\hat{\Sigma}$. Calling $\mathcal{I}$ the integral in eq.~(\ref{eq:n-entropy}), we have:
\begin{equation}
 \mathcal{I} = \frac{1}{Z_d(\hat{\Sigma},\nu)^{n+1}} \int \ud^{d} x\;   \left[  1+ \sum_{i=1}^{d} \left(\frac{1}{\lambda_i\nu}\right)x_{i}^2   \right]^{-\frac{1}{2}(n+1)(\nu+d)}
\end{equation}
Where $\lambda_{i}$ is the eigenvalue of $\Sigma$ corresponding to the $i$-th direction. We can also choose variables $r_{i} = \sqrt{\left(\frac{1}{\lambda_i\nu}\right)}x_{i}$ to write:
\begin{equation}
 \mathcal{I} = \Sd\frac{\sqrt{\nu^d|\hat{\Sigma}|}}{Z_d(\hat{\Sigma},\nu)^{n+1}}  \int_{0}^{\infty} \ud r\; r^{d-1}   \left[  1+ r^2 \right]^{-\frac{1}{2}(n+1)(\nu+d)}.
\end{equation}
The above integral is related to the Beta function yielding:
\begin{equation}
 \mathcal{I} = \Sd\frac{\sqrt{\nu^d|\hat{\Sigma}|}}{2Z_d(\hat{\Sigma},\nu)^{n+1}} B\left(\frac{1}{2} n (\nu+d) + \frac{\nu}{2} , \frac{d}{2}\right).
\end{equation}
Plugging it into  eq.~(\ref{eq:n-entropy}) and using our definition (\ref{eq:partitionfunction}) for $Z_d(\hat{\Sigma},\nu)$  gives:

\begin{eqnarray}
\label{eq:ddimentropy}
 S[p_d]  &=& \frac{1}{2}\log\left[{(\pi\nu)^d |\hat{\Sigma}|}\right] \\
&+& \log \frac{B\left(\frac{\nu}{2},\frac{d}{2}\right)}{\Gamma\left(\frac{d}{2}\right)} \nonumber \\
&+& \left(\frac{\nu+d}{2}\right) \left[\psi\left(\frac{\nu+d}{2}\right) - \psi\left(\frac{\nu}{2}\right) \right] \nonumber,
\end{eqnarray}

where $\psi\left(x\right)$ is the digamma function.

The mutual information of the Student d-dimensional distribution can be calculated using eq.~(\ref{eq:multimutual}) with the entropy given by eq.~(\ref{eq:ddimentropy}):
\begin{eqnarray}
I_d(\hat{\Sigma},\nu) &=& - \frac{1}{2}\log\mid \hat{\Sigma}\mid \\
&+&\log\left\lbrace\dfrac{\left[B\left(\frac{\nu}{2},\frac{1}{2}\right)\right]^d\Gamma\left(\frac{d}{2}\right)}{\pi^{\frac{d}{2}}B\left(\frac{\nu}{2},\frac{d}{2}\right)}\right\rbrace  - \frac{\nu(d-1)}{2}\psi\left(\frac{\nu}{2}\right)\nonumber \\
&+&\frac{d(\nu+1)}{2}\psi\left(\frac{\nu + 1 }{2}\right) - \frac{(\nu+d)}{2}\psi\left(\frac{\nu + d }{2}\right)\nonumber
\end{eqnarray}

Notice that the only term depending on the correlation matrix $\hat{\Sigma}$  is  the mutual information of a Gaussian distribution $I_{\mbox{Gauss}}=-\frac{1}{2}\log|\hat{\Sigma}|$. The remaining term is the information  excess. For $d=2$ we have:

\begin{equation}
I_2(\rho,\nu) =  I_{\mbox{Gauss}} + I_{\mbox{Excess}},
\end{equation}
with
\begin{equation}
I_{\mbox{Gauss}} = -\frac{1}{2}\log\left(1-\rho^2\right)
\end{equation}
and
\begin{eqnarray}
 I_{\mbox{Excess}} &=& 2\log\left(\sqrt{\frac{\nu}{2\pi}}B\left(\frac{\nu}{2},\frac{1}{2}\right) \right)  \\
                   &-& \frac{2+\nu}{\nu} + (1+\nu)\left[ \psi \left(\frac{\nu+1}{2}\right) - \psi \left(\frac{\nu}{2}\right) \right],\nonumber
\end{eqnarray}
where we used that $B(x,1) = \frac{1}{x}$ and  $\psi(x+1) - \psi(x) = \frac{1}{x}$.

\end{document}